# Activation and Alignment:
# A Causal Account of the Scientific Revolution

Harry Sticker
Ganymede Technology
New York, New York, USA
hsticker@ganymedetechnology.com

**Abstract:**

Standard historiographical approaches to the Scientific Revolution illuminate background conditions but leave three puzzles unresolved: what triggered the initial escalation of inherited tensions, what made early investigative efforts durable, and why natural philosophy became the locus of transformation rather than theology, law, or classical scholarship. This paper develops a causal account by identifying the mechanisms of activation at the individual level and the institutional alignment that converted rare psychological drive into durable research traditions.

The trigger architecture operates at two levels. At the individual level, activation occurs when investigators experience inherited puzzles as psychologically intolerable; capture stabilizes inquiry through cognitive, material, and social entanglements; and externalization converts methods into transmissible forms. At the institutional level, role expansion embeds elevated standards into positions; succession ratchets prevent regression through competitive selection; and domain channeling directs institutional energy toward particular fields.

A systematic comparison across Islamic, Chinese, and European cases demonstrates that each component is necessary, but none is sufficient on its own. The Scientific Revolution occurred when all components aligned at Padua-Venice and Oxford-London, where corporate autonomy, competitive appointments, and state patronage converged with investigative practices. The Galileo case provides decisive evidence: his selective activation across domains demonstrates that activation operates as a specific mechanism rather than a stable dispositional trait.

**KEYWORDS**: Scientific Revolution, philosophy of science, scientific change, institutional mechanisms, psychology of inquiry, comparative civilizations, succession ratchets

# 1. Introduction: The Puzzles of Divergent Outcomes

For more than fifteen hundred years, every component necessary to construct a thermometer existed. The pneumatic principles described by Hero of Alexandria in the first century CE provided all the necessary apparatus. Glassblowing techniques allowed the fabrication of narrow tubes and bulbs. The thermal expansion of air and liquids had been observed since antiquity. Numerical scales for dividing continuous quantities were standard in astronomy, geography, and commerce. Yet no one built a thermometer. Even when Galileo constructed a thermoscope around 1593, he treated it as a demonstration device—a curiosity that showed thermal expansion—rather than as a precision instrument for measurement. Around 1610, at the University of Padua, Santorio Santorio, a professor and well-respected physician, adapted Galileo's device into a calibrated medical thermometer. He added a numerical scale, standardized the construction, and used it systematically to quantify fevers and metabolic changes. Within a few decades, thermometers became standard instruments across Europe (Middleton 1966). The fifteen-hundred-year delay cannot be explained by the absence of materials, techniques, or knowledge. Even the seventeen-year gap between Galileo's thermoscope and Santorio's thermometer cannot be explained by technical constraints—the transformation required only the addition of a scale.

What changed was not technological capacity but investigative urgency: Santorio experienced the absence of quantitative precision as something that had to be overcome. The thermometer represents a case of precision seeking a problem—rising instrumental standards eventually forced reconceptualization of disease itself (Bigotti 2020).

This pattern—invention without escalation, followed by transformation through obsessive pursuit—recurs throughout the early modern period. In 1609 and 1610, Thomas Harriot and Galileo Galilei observed the Moon and Jupiter's moons with telescopes of similar construction, possessed comparable mathematical training, and faced identical observational opportunities. Yet their responses differed sharply. Galileo attacked the opportunity relentlessly. He rapidly improved the telescope, pursued nightly observations under difficult conditions, drew elaborate diagrams of lunar topography and Jupiter's satellites, and, within weeks, had written, edited, and arranged the publication of his *Sidereus Nuncius*. He spent months defending his discoveries against skeptics and integrated them into his cosmological perspective (Galilei 1610). By contrast, Harriot recorded his observations carefully in notebooks, shared them with correspondents, and then set them aside. As a known Copernican, he likely understood the cosmological significance, but he published nothing and made no effort to develop his observations into a sustained program of exploration, even though he was an explorer (Shirley 1983; Bucciantini et al. 2015). The contrast is not about instruments, data, or training. It is about what triggered one investigator to escalate a puzzle into a long-term inquiry while another, equally equipped, did not.

The thermometer and telescopic cases reveal the same underlying pattern: activation is independent of invention or initial observation. Galileo invented the thermoscope, but did not pursue the precision necessary to make it a useful device; Santorio did. Harriot observed Jupiter's moons but did not experience their implications as demanding immediate pursuit; Galileo did. In both cases, the technical capacity existed, the opportunity was present, and the initial step had been taken. What differed was the threshold at which imprecision or unexplored implications became psychologically unbearable.

These cases are not isolated examples. Early modern Europe produced many short-lived bursts of technical or conceptual innovation, yet only a few crystallized into stable research programs. Galileo's telescopic astronomy persisted, expanded, and generated successors across generations. The question is not simply why some ideas were correct or useful, but why certain investigative efforts acquired the cognitive, social, and institutional reinforcement needed to survive their fragile early stages while others, equally promising, failed to stabilize.

Even if we grant that early modern Europe was unusually fertile ground for innovation, the question remains: why did the most explosive and self-sustaining transformation center on natural philosophy rather than on other domains with equally deep traditions and pressing puzzles? Law, medicine, theology, and classical philology all possessed long-standing tensions, rich textual corpora, skilled practitioners, and institutional support. Medical faculties at Padua, Bologna, and Montpellier had been operating for centuries. Legal scholarship had generated sophisticated interpretive traditions. Theological disputation had reached extraordinary levels of refinement in the scholastic universities. Yet it was natural philosophy that became the locus of a systematic, escalating, and ultimately transformative investigative enterprise.

Taken together, these cases reveal a common explanatory deficit. Comparable conditions repeatedly produced divergent outcomes: some investigators escalated puzzles, while others did not; some lines of inquiry stabilized, while others evaporated; some domains became sites of explosive innovation, while others remained unchanged. Three puzzles emerge:

**The activation puzzle**: What triggers the initial escalation? What determines when an investigator experiences an inherited tension as an intolerable demand for resolution rather than a background imperfection to be accommodated or deferred?

**The durability puzzle**: What sustains investigation long enough for it to become a durable tradition? What prevents early innovations from dissipating when obstacles accumulate or competing demands intervene?

**The direction puzzle**: Why did natural philosophy become the primary target in Europe? What channeled investigative energy, institutional support, and rising standards toward natural philosophy rather than toward theology, law, medicine, or classical scholarship?

When the Scientific Revolution occurred in Europe during the sixteenth and seventeenth centuries, it broke a two-thousand-year pattern. Puzzles that had long been accommodated by scholars and philosophers were suddenly attacked with new intensity. This seemingly sudden transformation has elicited extensive scholarly attention. Historians and philosophers of science have illuminated its conceptual shifts (Koyré 1957; Kuhn 2012), its social contexts (Shapin 1996; Biagioli 1993), its technological preconditions (Eisenstein 1980; Bennett 1987), its material practices (Shapin and Schaffer 2011), and its institutional settings (Dear, 2006; Gaukroger, 2006). These works give us a deep understanding of what happened, the background conditions, and why it mattered. Steven Shapin captured the resulting interpretive complexity with a famous paradox: "There was no such thing as the Scientific Revolution, and this is a book about it" (Shapin, 1996, p. 1). His point was not to deny transformation but to challenge singular explanations. The Scientific Revolution was not a unified event but a convergence of multiple processes.

Yet even this rich historiographical tradition leaves the three puzzles unresolved. Identifying key figures—Galileo, Kepler, Descartes, Boyle, Newton—does not explain why these investigators undertook arduous, often unrewarded efforts while others with comparable skills and opportunities did not. Documenting conceptual shifts does not explain why certain tensions became actionable at particular moments. Analyzing institutional contexts does not explain why some innovations became embedded as permanent standards while others remained optional. Tracing technological developments does not explain the thermometer's 1600-year delay or why instruments often followed investigative urgency rather than enabling it. Some accounts analyze experimental routines, instruments, and representational conventions to illuminate how investigators engaged problems (Dear 2006; Shapin and Schaffer 2011). Yet even these accounts do not identify the micro-processes that made particular tensions actionable for particular investigators at particular times.

What is missing is a trigger architecture: a structured account of the processes that converted latent possibilities into a self-propelling investigative enterprise. Standard historiographical tools—conceptual analysis, institutional history, material culture studies, biographies—illuminate the landscape in which the Scientific Revolution occurred. But they do not explain the ignition sequence. They do not tell us why the thermometer was finally built around 1600, why Galileo escalated while Harriot did not, or why natural philosophy rather than theology became the site of rising standards. To resolve these puzzles, we must identify the mechanisms that:

- converted inherited tensions into urgent investigative projects (activation),
- stabilized fragile early inquiries into sustained programs (durability), and
- directed institutional energy toward natural philosophy rather than other domains (direction).

## 2. Rehabilitating Psychology in the Philosophy of Science

Explanations of the Scientific Revolution have traditionally avoided psychological factors. Logical positivists dismissed them as psychologism and relegated mental processes to the "context of discovery," rendering them methodologically irrelevant (Carnap 2003; Reichenbach 1938). The division was sharp: discovery belonged to psychology, biography, and contingency; justification belonged to logic, evidence, and rational reconstruction. Only the latter was considered philosophically legitimate. By mid-century, models of scientific change — Kuhn's paradigms and Lakatos's research programmes—shifted attention to conceptual and social structures while explicitly excluding individual-level psychological variations (Kuhn 2012; Lakatos 1970). Even when Kuhn introduced psychological elements such as gestalt shifts and conversion experiences, these remained descriptions of general cognitive states rather than explanatory mechanisms. Practice-based and sociological approaches, which emphasized experiment, instrumentation, and laboratory culture, continued to treat psychology as illegitimate for explanation (Hacking 1983; Latour and Woolgar 1986). The operating assumption across these traditions was that psychological variables — temperament, motivation, curiosity — were too idiosyncratic, too personal, and too resistant to systematic analysis to carry explanatory weight in the history and philosophy of science.

This exclusion left critical explanatory gaps. When external conditions are similar but individual responses differ, some internal threshold must account for the difference. The challenge is to operationalize this threshold without collapsing into biographical speculation.

Recent philosophical work has reopened the space of individual psychological explanations by showing that scientific reasoning depends on cognitive and psychological processes that cannot be reduced to formal logic or social structure (Giere 2006; Morrison 2000, 2015; Nersessian 1995). Model-based reasoning, analogical thinking, and conceptual blending all involve psychological mechanisms that shape how investigators formulate problems, evaluate solutions, and decide when to pursue or abandon lines of inquiry. These mechanisms are not merely heuristics or aids to discovery; they are the essence of scientific reasoning. The account developed here advances this trajectory by adding a specific, operationalizable psychological mechanism that helps explain why some investigators escalated inherited puzzles while others, facing the same conditions, did not.

The psychological factor reintroduced here is not the usual suspects of genius or curiosity, nor a dispositional trait that some possess and others lack. It is a specific psychological mechanism: a threshold of intolerance for unresolved tensions that emerges through a specific sequence of inquiry, not from pre-existing traits. Dispositional factors may explain baseline differences in investigative persistence—why some individuals are more likely than others to interrogate problems deeply. But dispositional explanations alone are insufficient: they do not explain why the same investigator activates selectively across different domains, nor do they specify the mechanism by which interrogation generates intolerance for unresolved tensions. The intolerance is generated by a sequence that begins with intensive interrogation of known puzzles, followed by the perception of previously invisible gaps, leading to a psychological compulsion to resolve them. The intolerance is not present before the interrogation begins; it is created through the recursive process of the interrogation itself, and by pursuing a question beyond normal stopping points, the investigator constructs a psychological trap from which disengagement becomes increasingly difficult. This sequence can be identified in the historical record through behavioral markers: the language investigators use to describe problems, the levels of precision they pursue, the duration of their engagement, and the recurrence of similar responses across different problem domains. These markers are not retrospective clinical diagnoses; they are observable outputs of the interrogation process.

This distinction between dispositional traits and context-emergent mechanisms aligns with recent work in personality psychology. Traits are stable, enduring characteristics that manifest consistently across contexts; states are temporary conditions that emerge from specific situational features (Chaplin et al. 1988). Whole Trait Theory demonstrates that even stable trait patterns emerge from distributions of context-dependent states rather than from fixed dispositions, with social-cognitive mechanisms—including goal structures, interpretations of significance, and situational appraisals—determining which behaviors are enacted in specific contexts (Fleeson and Jayawickreme 2015). Activation operates as a state-like mechanism in this framework: it emerges from how specific problems are interrogated in specific contexts, mediated by the investigator's perception of gap severity, tractability, and significance. Dispositional factors—such as general intellectual persistence, tolerance for ambiguity, or propensity for deep engagement—may predispose some investigators toward the intensive interrogation that triggers activation. But disposition alone cannot explain within-individual selectivity: why Galileo relentlessly pursued telescopic astronomy but not thermometry, why Kepler relentlessly pursued planetary theory but contemporary astronomers with similar training did not, or why Santorio relentlessly pursued metabolic quantification while Galileo, working at the same institution with the same device, treated the thermoscope as a curiosity. These patterns demonstrate that activation depends on the interaction between investigator characteristics, problem structure, and interrogation depth, not merely on stable traits carried uniformly across contexts.

Reintroducing psychology in this way is not a return to internalism. Internalist historiography explained scientific change through the autonomous development of ideas, treating external factors as mere context (Koyré 1957). The account here, by contrast, treats psychological thresholds as embedded in a larger context of material practices, institutional structures, and social configurations. Activation initiates escalation, but it does not determine outcomes. Whether an activated inquiry becomes a durable tradition depends on capture mechanisms (cognitive, material, and social entanglements) and institutional scaffolding (role expansion, succession ratchets, and domain channeling). Psychology enters the explanation not as a sufficient cause but as a necessary initiating condition.

## 3. The Trigger Architecture

This architecture integrates individual and institutional mechanisms. Activation, capture, and externalization describe how a single investigator's work accumulates and stabilizes. Role expansion, succession ratchets, and domain channeling determine whether that individual achievement persists across generations and becomes a durable research tradition. The relationship is hierarchical: individual mechanisms provide the ignition, institutional mechanisms provide the oxygen. Without ignition, nothing starts; without oxygen, the fire dies. These mechanisms are interdependent: individual processes initiate investigative trajectories, but only institutional structures render them historically consequential.

The individual-level process unfolds through three stages. Activation identifies the psychological threshold at which an inherited puzzle becomes unavoidable. Capture explains how an initially fragile investigative impulse becomes stabilized through cognitive, material, and social entanglements. Externalization accounts for how captured inquiry becomes transmissible: methods, instruments, and problem-structures are converted into forms that others can adopt without sharing the originating investigator's psychological threshold. Together, these stages provide the causal structure needed to resolve the activation and durability puzzles.

## 3.1. Activation

The first stage of the trigger architecture is activation, the moment when an inherited puzzle demands resolution rather than remaining something merely noted, accommodated, or deferred. Activation is selective. Many early modern investigators possessed the skills, tools, and opportunities to address the same tensions, yet only a few experienced them as intolerable anomalies that demanded sustained inquiry. The thermometer case exemplifies this selectivity: every component had existed for 1,500 years, and even when Galileo constructed a thermoscope around 1593, he treated it as a demonstration device rather than a precision instrument. Only Santorio, around 1610, experienced the absence of quantitative temperature measurement as intolerable and transformed Galileo's device into a calibrated thermometer. The Harriot-Galileo telescopic contrast shows the same pattern: identical observational opportunities, radically different responses. Activation, therefore, requires a threshold condition—one that determines when a puzzle crosses from background noise into something that demands sustained investigation.

This threshold is historically contingent. For most of antiquity and the Middle Ages, the dominant epistemological framework in the mathematical sciences was the doctrine of saving the appearances (sōzein ta phainomena). The task of theory was to reproduce observed phenomena as accurately as possible, not to disclose the physical or causal structure responsible for them. Within this framework, predictive adequacy was sufficient; deeper questions about physical reality or underlying causes were bracketed as philosophically optional or methodologically premature. From Ptolemy onward, astronomical models were evaluated primarily by their success in matching observations, and multiple incompatible constructions could be tolerated so long as they delivered the same predictions. This behavior was not intellectual laziness or conceptual confusion, but a well-entrenched professional standard governing what counted as an adequate explanation (Kuhn 1976; Lloyd 1989; Swerdlow and Neugebauer 2012). Against this backdrop, the refusal to accept empirical adequacy without deeper coherence marks a departure from normal practice rather than its continuation. The Scientific Revolution can thus be understood as a collapse of inherited adequacy thresholds: what 14th-century scholars found tolerable—multiple incompatible models, qualitative description without measurement, observation without systematic pursuit—became intolerable for certain 17th-century investigators.

Activation occurs when an investigator encounters a tension that existing frameworks cannot absorb, and that becomes psychologically intolerable. Such intolerance is not reducible to temperament or genius; it emerges from a specific micro-sequence of inquiry. Driven investigators engage in obsessive interrogation of a problem at depths others find unnecessary. Through this interrogation, they perceive inconsistencies or gaps that were previously not salient. Once perceived, these gaps become compulsions: the investigator experiences them as demands for resolution. In some cases, the investigator deepens a puzzle through interrogation that others would have left at a superficial level. The investigator constructs the trap through the depth of engagement.

### Identifying Activation Through Contrast Pairs

Because activation is not directly observable as a mental state, it must be inferred from behavior. The most reliable method is the use of contrast pairs: comparing investigators who faced similar external conditions—data, tools, training, institutional position—but produced radically different investigative outcomes. When external factors are matched, and divergence persists, the difference isolates an internal threshold as the causal variable. This method avoids speculative psychologizing by grounding claims in documented behavior rather than inferred mental states.

| Domain | Investigator A (Not Activated) | Investigator B (Activated) | Key Divergence |
|---|---|---|---|
| Thermometry | Galileo (thermoscope as curiosity) | Santorio (calibrated thermometer) | Qualitative demonstration vs. quantitative precision |
| Telescopic Astronomy | Harriot (observations recorded) | Galileo (sustained cosmological program) | Interesting data vs. intolerable cosmological necessity |
| Planetary Theory | Ptolemaic tradition (saving appearances) | Copernicus/Kepler (physical coherence) | Predictive adequacy vs. elimination of mathematical "monstrosities." |

*Table 1: The Activation Case*

Two cases exemplify this method. The Harriot-Galileo telescopic contrast provides the first example. Both observed the Moon and Jupiter's moons in 1609-1610 with comparable telescopes. Both possessed mathematical training. Both had access to similar observational opportunities. Yet Galileo escalated the observations into a sustained program—improving instruments, producing systematic diagrams, publishing rapidly, defending discoveries publicly—while Harriot recorded observations, shared them privately, and moved on. Harriot understood the cosmological significance—he was a known Copernican—but recorded his observations without pursuing them into a sustained program (Shirley 1983). But understanding significance is not the same as experiencing it as intolerable. Harriot felt no compulsion to resolve the tension; Galileo did.

The Galileo-Santorio thermometer contrast provides the second example. Galileo constructed a thermoscope around 1593, but treated it as a demonstration device. Working at the same institution, Santorio took Galileo's device and experienced its lack of quantitative precision as intolerable. He added numerical scales, standardized construction, and pursued a thirty-year program of metabolic quantification (Bigotti 2020; Middleton 1966). Both occupied prestigious medical positions at Padua. Both had access to the same device. Yet Santorio experienced the lack of measurement precision as a gap that demanded resolution, while Galileo—the device's inventor—did not.

These contrasts isolate activation. External conditions do not explain the divergences. In both cases, the difference lies in whether the puzzle crossed a psychological threshold—whether it became something that had to be pursued rather than something that could be pursued.

## Operational Markers of Activation

Activation leaves distinctive behavioral signatures that can be identified in the historical record. Three operational markers reveal when an investigator has crossed the activation threshold:

**Severity of Framing**

Driven investigators characterize gaps or inconsistencies as fundamental inadequacies requiring resolution, not as improvable imperfections. Their language signals necessity rather than preference. Galileo's correspondence and publications describe telescopic observations with urgent intensity: the implications for cosmology cannot be deferred or accommodated within existing frameworks; they demand immediate attention and integration (Galilei 1610; Drake 1978). Similarly, Copernicus framed Ptolemaic astronomy's use of equants and multiple incompatible models as intolerable incoherence rather than improvable imprecision—portraying the system as a monstrous assemblage of mismatched parts that violated fundamental principles of celestial motion (Westman 2011). Kepler used similarly visceral language to describe astronomical inconsistencies, framing the 8-arcminute discrepancy as a 'deception' perpetrated by the assumption of uniform circular motion (Stephenson 1994). By contrast, Harriot recorded telescopic observations comparable to Galileo's but treated them as interesting findings to be noted and shared, not as intolerable tensions requiring resolution. The difference is not in what Galileo and Harriot saw but in how each experienced the gap between observation and existing theory.

**Precision Beyond Utility**

Activated investigators pursue levels of precision that exceed any practical, institutional, or social adequacy threshold. The precision is epistemic, not instrumental.

The canonical case is Kepler's pursuit of Mars's orbital discrepancies: he spent years resolving an 8-minute arc discrepancy that had no consequences for navigation, calendric reform, or astronomical prediction at the practical level demanded by patrons or institutions. The discrepancy was invisible to nearly every contemporary astronomer, yet Kepler experienced it as intolerable. This obsessive precision strained his patronage relationships—Emperor Rudolf II expected the practical Rudolphine Tables, not years spent on Mars's orbit—and delayed more remunerative astronomical work (Voelkel 2001; Westman 2011).

Santorio's transformation of Galileo's thermoscope into a calibrated thermometer shows the same pattern. Medical diagnosis had proceeded for centuries through qualitative assessment of fever. Adding numerical scales and pursuing systematic quantification served no immediate therapeutic purpose—Galenic medicine required only qualitative assessment of fever intensity (Bigotti 2020). Yet Santorio found the lack of measurement precision intolerable. Similarly, Galileo's telescopic observations went far beyond what was necessary to document lunar topography or Jupiter's satellites: he pursued systematic observation, detailed diagrams, and instrumental refinement with an intensity that exceeded any external demand.

**Sustained Engagement**

Investigation persists for years or decades beyond institutional or economic incentives. The puzzle remains psychologically active until it is resolved or the investigative framework is rebuilt. Such patterns are well-documented: Copernicus worked on planetary theory for over three decades, revising and refining his heliocentric model long after the initial conceptual breakthrough (Westman 2011). Kepler pursued Mars's orbit for nearly a decade, through financial hardship and institutional instability. Galileo's telescopic program continued through opposition, institutional obstacles, and personal risk. The work was not episodic but continuous—evident in correspondence, unpublished notes, and successive refinements of instruments and observations. Santorio's metabolic quantification program lasted thirty years, far exceeding any external reward structure. This sustained engagement distinguishes activated inquiry from opportunistic investigation that terminates when obstacles accumulate or rewards diminish.

**Within-Individual Variation: The Decisive Test**

The two preceding contrasts demonstrate between-individual variation: different investigators responding differently to the same puzzle. But Galileo's case provides even stronger evidence: within-individual variation across different domains.

Galileo appears on both sides of the activation threshold. With telescopic astronomy, he escalated aggressively: rapid instrumental refinement, systematic observation, public defense, and decades of sustained engagement. With thermometry, he did not: he invented the thermoscope around 1593 but treated it as a qualitative demonstration device, never pursuing the quantitative precision that Santorio later imposed. Both opportunities were equally available to him; both occurred at the same institution during overlapping periods; both involved similar technical challenges (adding scales, standardizing construction). Yet one triggered activation, the other did not.

Dispositional traits may make activation more likely—investigators with high persistence or low tolerance for ambiguity may, on average, interrogate problems more deeply. But dispositional traits predict consistency across opportunities. If activation were purely a stable personality trait—if Galileo's investigative intensity were simply a dispositional constant—it would manifest uniformly across domains. A dispositional explanation predicts that the curious investigator pursues all puzzles; the driven investigator escalates all opportunities. But Galileo was selective. Something about telescopic observations triggered the recursive interrogation that generated psychological intolerance; something about thermal expansion did not.

Within-individual variation is more diagnostic than between-individual variation because it controls for all individual-level factors: training, institutional position, available resources, social networks, and dispositional traits. The only variable that differs is the puzzle itself and how Galileo interrogated it. This isolates activation as a content-dependent mechanism—one that emerges from how a specific puzzle is engaged—rather than as a trait the investigator carries uniformly across contexts. This pattern aligns with contemporary personality research, which finds that trait-relevant behaviors emerge from social-cognitive mechanisms operating on the features of specific situations rather than from stable dispositions alone (Fleeson & Jayawickreme, 2015). Activation is then better understood as a state trait that arises when deep interrogation of a specific problem reveals intolerable gaps, rather than as a general disposition that manifests identically across all investigative opportunities.

**Bundle Logic: Why Activation Is a Distinct Component**

Within-individual variation demonstrates that activation cannot be explained by dispositional traits alone, even when such traits may contribute to the likelihood of deep interrogation. But this raises a further methodological question: could the three markers themselves be explained by external factors rather than by a distinct psychological mechanism? By itself, each marker can be ambiguous. Severity of framing could be a rhetorical strategy; excessive precision could be technical virtuosity; persistence could be career ambition. But taken together, the markers form a bundle that cannot be explained by social, institutional, or instrumental factors alone.

Patronage explains compliance with standards, not obsessive excess beyond those standards. Kepler's pursuit of eight-minute precision exceeded what any patron demanded. Career ambition explains public work aimed at recognition, not decades of private investigation that remains unpublished. Copernicus's thirty-year refinement of *De revolutionibus* before publication does not fit a career-advancement model. Training explains competence within established methods, not deviation from contemporaries who are equally trained and do not exhibit the same investigative depth. Harriot had training equivalent to Galileo's but did not escalate observations into a sustained program.

Combinations of external factors fare no better. Patronage plus career ambition might explain sustained public work, but not Copernicus's three decades of private refinement before publication. Training and institutional opportunities might explain competent performance within a domain, but not Kepler's pursuit of precision, which actively damaged his patronage relationships. Even if we grant that Galileo possessed unusual opportunities—Paduan appointment, Medici patronage, instrument-making networks—these advantages were shared by colleagues who did not escalate comparable puzzles. The conjunction of markers resists reduction because it produces outcomes that run counter to institutional incentives rather than align with them.

The key insight is that no single external factor, nor any plausible combination of external factors, accounts for the simultaneous presence of all three markers. When an investigator:

- frames problems as fundamental inadequacies (not improvable imperfections),
- pursues precision far beyond practical thresholds, and
- sustains engagement across decades without external reward,

the pattern cannot be reduced to social opportunity, institutional incentive, or technical training. The conjunction isolates activation as a distinct causal component: a threshold of intolerance for unresolved tensions that transforms inherited puzzles into urgent investigative projects.

Activation is not an argument for psychological determinism. It does not guarantee outcomes. Many investigators who crossed the activation threshold produced no lasting impact because their work was not captured or institutionally supported. What the bundle logic establishes is that activation is a necessary initiating condition—one that cannot be explained away by external factors and must be integrated into historical explanations of why some investigators escalated puzzles while others did not.

Activation is therefore the first stage of the trigger architecture—the ignition point that sets the subsequent stages of capture and externalization into motion.

## 3.2. Capture: When Activated Inquiry Becomes Durable

Activation is necessary but not sufficient. Many investigators experience a moment of intolerable tension, pursue a line of inquiry with initial intensity, and then abandon it when obstacles accumulate or competing demands intervene. For activation to have historical consequence, it must be captured—drawn into structures that preserve, amplify, and stabilize the initial burst of inquiry. Capture is therefore the second stage of the trigger mechanism: the set of processes that prevent activated inquiry from dissipating and instead convert it into a durable investigative trajectory. Like other practice-based accounts of scientific stability, capture operates through the cognitive, material, and social routines that shape how investigators engage problems.

Capture operates through three interlocking channels: cognitive, material, and social. At the cognitive level, early investigative steps must generate further tensions rather than closure. A successful measurement, observation, or calculation does not resolve the initial puzzle; it exposes new inconsistencies or demands for refinement. This recursive generation of puzzles keeps the investigator psychologically locked into the project. At the material level, instruments, diagrams, tables, and notebooks create external scaffolding that stores partial results and makes abandonment costly. Once an investigator has built an apparatus, accumulated data, or constructed representational tools, the work acquires inertia: it becomes easier to continue than to discard. At the social level, even minimal recognition—an encouraging letter, a skeptical query, a request for clarification—creates obligations that tether the investigator to the project. These forms of capture need not be institutionalized; they can be as small as a correspondent asking for an update or a student requesting a demonstration. Capture is therefore not mere momentum; it is the creation of entanglements that make disengagement increasingly costly.

**Galileo's Telescopic Program: A Worked Example**

Galileo's telescopic work provides the clearest demonstration of how all three channels of capture operate together over time. The sequence shows how early observations generated recursive puzzles, how material commitments accumulated, and how social entanglements made withdrawal increasingly difficult.

*Cognitive capture through recursive puzzle generation*: Galileo's initial telescopic observations of the Moon in late 1609 revealed irregular surface features inconsistent with Aristotelian perfect spheres. Rather than resolving the puzzle, this discovery generated new questions. If the Moon had mountains, how high were they? What did this imply for the nature of celestial matter? When he turned the telescope to Jupiter in January 1610, he discovered four satellites, which raised further puzzles about orbital mechanics and the plurality of centers of motion. Each answer opened new problems. The phases of Venus, discovered in late 1610, required integrating them with the Medicean satellites and lunar topography into a coherent cosmological framework. By 1611, the discovery of sunspots added another layer: if the Sun itself was mutable, what remained of celestial perfection? Each discovery did not close an investigation but compelled further refinement and integration (Galilei 1610; Drake 1978; Bucciantini et al. 2015).

*Material capture through apparatus and documentation*: Galileo's early telescopic successes required sustained investment in materials. He improved the telescope from 8x to 20x magnification between November 1609 and March 1610, requiring iterative lens-grinding and tube construction (Bucciantini et al. 2015; Van Helden 1977). He produced detailed drawings of the Moon's surface, which became representational commitments: once created, they demanded justification, defense, and refinement. His tables of Jupiter's satellite positions accumulated over months, creating a dataset that represented sunk effort. Abandoning the telescopic program would have meant discarding these material and intellectual investments. The apparatus itself became a stabilizing force: having built better telescopes, Galileo was positioned to make further discoveries that others could not replicate  (Bucciantini et al. 2015; Reeves 2008).

Social capture through correspondence and controversy: Even before publishing *Sidereus Nuncius*, Galileo received encouragement from Venetian patricians and early telescope viewers. After publication in March 1610, he entered correspondence with Kepler, who wrote enthusiastically in *Dissertatio cum Nuncio Sidereo* (1610). Galileo's social behavior created an expectation: Kepler and others anticipated further discoveries and refinements.

Skeptics such as Martin Horky and Giulio Libri challenged his observations, which forced Galileo to defend and extend his claims. Each controversy created an obligation to respond. When Galileo moved to Florence as court mathematician to the Medici in 1610, the position itself created expectations of continued celestial discoveries. The social network around his telescopic work—correspondents requesting updates, skeptics demanding proof, students seeking instruction—tethered him to the program (Bucciantini et al. 2015; Biagioli 1993).

This worked example shows how capture is not a single event but an accumulating process. Each channel reinforced the others: cognitive puzzles required material refinement, which generated social attention, which created obligations to pursue further cognitive puzzles. The publication of Sidereus Nuncius in March 1610 exemplifies how capture becomes irreversible: once Galileo publicly claimed to have discovered four new celestial bodies and mountains on the Moon, he was intellectually and socially entangled (Bucciantini et al. 2015; Biagioli 1993). He could no longer retreat to earlier positions without losing his newly won status as Florentine court mathematician. The psychological trap he constructed for himself became a public commitment that the world recognized and demanded he defend. By mid-1610, Galileo was captured. Withdrawal would have been cognitively unsatisfying (puzzles would remain unresolved), materially wasteful (the apparatus and data would be discarded), and socially costly (expectations would be violated). Capture converted his initial activation into a self-reinforcing investigative trajectory.

This worked example demonstrates how capture operates through interlocking channels that reinforce one another. The telescopic program shows all three mechanisms—cognitive, material, and social—operating simultaneously to create a self-sustaining investigative dynamic. These mechanisms can be identified through specific operational markers.

**Operational Markers of Capture**

Because capture is not directly observable as a mental state, it must be inferred from patterns of entanglement. Three operational markers identify when an activated inquiry has been captured:

**Recursive Puzzle Generation**

Early results produce new tensions that demand further work. The project expands rather than contracts, distinguishing captured inquiry from projects that terminate once an initial goal is achieved.

**Material Inertia**

The investigator constructs or refines apparatus, tables, diagrams, or representational tools that store partial results and make abandonment costly. Material scaffolding becomes a stabilizing force.

**Social Tethering**

Correspondence, requests, challenges, or even opposition create obligations that keep the investigator engaged. These tethers can be minimal; what matters is that they generate expectations the investigator feels compelled to meet.

Each marker alone is ambiguous. Recursive puzzles could reflect methodological difficulty; apparatus construction could reflect technical virtuosity; social engagement could reflect patronage or collegiality. But taken together, they form a bundle that cannot be explained solely by external incentives. The conjunction of markers isolates capture as a distinct mechanism: the process by which activated inquiry becomes embedded in a reinforcing environment that makes disengagement increasingly unlikely.

Capture is therefore the bridge between activation and externalization. Activation explains why an investigator begins; capture explains why the investigator continues long enough for the work to accumulate, stabilize, and become transmissible. Capture creates the conditions under which an inquiry can acquire enough structure to be taken up by others. Without capture, activation produces only momentary intensity. With capture, it produces the extended, recursive labor that makes externalization possible.

### 3.3. Externalization: When Captured Inquiry Becomes Transmissible

Capture stabilizes an investigator's engagement, but stability alone does not produce historical consequence. Many captured inquiries remain idiosyncratic, isolated, or short-lived. For an investigative trajectory to endure—capable of generating successors, attracting resources, and reshaping disciplinary expectations—it must be externalized. Externalization is the third stage of the trigger architecture: the process by which captured inquiry becomes transmissible and independent of the originating investigator's psychological threshold.

Externalization occurs when the products of captured inquiry—methods, instruments, diagrams, problem-structures, and conceptual distinctions—acquire a life independent of the originating investigator. An externalized trajectory no longer depends on the originating investigator's intolerance for unresolved tensions; it persists because the tools and problem structures the investigator created enable others to continue the work.

Externalization, therefore, converts a personal investigative compulsion into a publicly available research program.

Externalization operates through three channels: methodological, conceptual, and material. Each channel makes different aspects of an investigation available for adoption by others who do not share the originator's activation threshold.

**Methodological Externalization**

Captured inquiry generates tools that encode the originating investigator's standards of adequacy. These may be mathematical procedures, observational protocols, representational conventions, or experimental scripts. Once externalized, these tools allow others to reproduce the investigator's results without sharing the investigator's psychological threshold.

Kepler's area law (equal areas swept out in equal times) embedded his demand for mathematical-physical coherence into a calculational procedure. Later astronomers could use the area law to compute planetary positions without sharing Kepler's intolerance for the Ptolemaic equant. The method externalized the standard (Stephenson 1994; Westman 2011).

Galileo's telescopic protocols—how to grind lenses, construct tubes, observe systematically, record positions—encoded his elevated observational standards into repeatable procedures. Other investigators could follow the protocols and achieve comparable results without experiencing Galileo's initial activation. The protocols became cognitive prostheses that allowed others to replicate his standards without sharing his psychological thresholds (Bucciantini et al. 2015).

Boyle's experimental recipes in *New Experiments Physico-Mechanical* (1660) provided scripts for replicating air-pump demonstrations. The recipes specified apparatus construction, procedural steps, and expected outcomes. By following the recipes, investigators could produce Boylean results without sharing Boyle's investigative drive. The recipes externalized experimental standards (Shapin and Schaffer 2011).

Methodological externalization transforms a private intolerance into a procedural requirement. Once methods are externalized, they circulate independently and become resources for further investigation.

**Conceptual Externalization**

Externalization occurs when an investigator's idiosyncratic puzzle becomes a generalizable problem-structure. A problem is conceptually externalized when its solution requires others to adopt the originating investigator's representational framework.

Kepler's reformulation of planetary motion as a geometric problem constrained by physical forces forced subsequent researchers to work within a narrower range of possibilities. Planetary theory could no longer be purely kinematic; it had to account for physical causes. This reformatting channeled subsequent investigators—Borelli, Hooke, Newton—into a shared trajectory. The problem-structure itself made certain approaches natural and others costly (Westman 2011; Wilson 1989).

Galileo's kinematic analyses redefined what counted as an adequate account of motion. Motion could no longer be explained through qualitative tendencies; it required quantitative relationships between distance, time, and acceleration. This reformatting made pre-Galilean accounts of motion seem inadequate even to investigators who had not experienced Galileo's initial activation (Drake 1978).

Descartes's analytic geometry restructured what it meant to pose a mathematical question. Geometric problems were reformulated as algebraic equations; solutions required symbolic manipulation. This reformatting created a new problem-space that successors entered as a matter of course (Bos 2001; Mancosu 2008).

Conceptual externalization creates a funnel: it channels subsequent investigators into a shared trajectory, making deviation costly and continuation natural. The problem-structure becomes self-propagating.

**Material Externalization**

Externalization requires that the products of captured inquiry become embedded in material and social structures that stabilize their circulation. These structures may include correspondence networks, instrument-makers' workshops, print culture, pedagogical routines, or patronage systems. Material externalization does not require formal institutions; early modern Europe's loose networks were sufficient. What matters is that the products of inquiry generate expectations, obligations, and opportunities for further work.

Galileo's telescopes were reproduced by instrument-makers across Europe, creating a material infrastructure that outlasted his personal involvement. Kepler's tables were incorporated into almanacs and navigation manuals, making them resources for practitioners who never engaged with his theoretical commitments. Boyle's air pumps were constructed in multiple locations, embedding his experimental program into geographically dispersed sites (Bennett 1987; Bucciantini et al. 2015).

Once materially externalized, a research trajectory becomes self-propelling: new entrants encounter a landscape already shaped by the originating investigator's commitments. The trajectory persists not because successors share the originating investigator's psychology, but because the tools, problem structures, and material infrastructures make continuation easier than starting anew.

**Why Individual-Level Externalization Is Not Sufficient**

Externalization at the individual level explains why an investigator's work becomes methodologically transmissible and attracts followers. But transmissibility alone does not guarantee persistence across generations. Many promising methods were adopted briefly and then abandoned; many innovations remained optional rather than obligatory. For investigative standards to rise irreversibly and across generations, individual-level externalization must be reinforced by institutional mechanisms that preserve and enforce elevated standards.

Galileo's telescopic program externalized at the individual level: his methods, instruments, and problem-structures circulated widely. But whether this trajectory persisted depended on institutional mechanisms—role expansion, succession ratchets, and domain channeling—that are analyzed in Section 4. Individual externalization creates the possibility of persistence; institutional mechanisms make that persistence durable and irreversible.

Externalization is therefore the point at which an investigative trajectory becomes capable of independent propagation. Activation explains why an investigator begins; capture explains why they continue; externalization explains why their work can outlast them and attract successors. Without externalization, captured inquiry remains a personal obsession. With externalization, it becomes a transmissible program—capable of generating followers, attracting resources, and reshaping the domain in which it operates.

# 4. Institutional Mechanisms: From Individual Achievement to Collective Obligation

The individual-level process—activation, capture, and externalization—explains how a single investigator's work accumulates, stabilizes, and becomes transmissible. But individual-level externalization alone does not produce historical transformation. For investigative trajectories to endure across generations, become obligatory rather than optional, and reshape entire domains, they must be reinforced by institutional mechanisms that preserve and enforce elevated standards.

Copernicus's heliocentric model exemplifies this pattern: it circulated widely as a calculational tool, was adopted by skilled astronomers, and influenced cosmological thinking. Yet it remained optional rather than obligatory. Universities continued to teach

Ptolemaic astronomy; no institutional penalty accompanied the rejection of heliocentrism. The difference between trajectories that persisted and those that remained optional lies not in individual-level externalization but in whether institutional mechanisms captured and preserved them.

The distinction between individual and institutional levels is critical. At the individual level, externalization occurs when an investigator's methods and problem structures become transmissible. Galileo's telescopic protocols, Kepler's area law, and Boyle's experimental recipes all externalized in this sense: they circulated as tools that others could adopt. But adoption remained voluntary. An astronomer could choose to ignore Kepler's area law; a natural philosopher could decline to replicate Boyle's experiments. Individual-level externalization creates the possibility of uptake but does not compel it.

Institutional mechanisms, by contrast, convert voluntary adoption into structural obligation. When role expansion is in effect, an individual's elevated practice becomes the minimum standard for occupying a position. When succession ratchets operate, competitive selection ensures that those standards cannot be lowered without professional penalty. When domain channeling operates, prestige gradients and structural affordances direct institutional energy toward particular fields, making them the sites where rising standards accumulate. These mechanisms do not depend on successors sharing the originating investigator's psychological threshold. They channel universal motivations—employment, status, recognition—toward maintaining and elevating standards.

What follows analyzes each mechanism in turn: how it operates, what structural conditions it requires, and how it transforms individual achievement into institutional permanence.

### 4.1. Role Expansion

Role expansion is the first institutional mechanism that converts individual investigative drive into durable professional standards. It refers to an institution's capacity to embed rising standards directly into the definition of a position, without requiring external authorization. When role expansion is present, a single investigator's elevated practice can become a mandatory requirement for all future incumbents. When absent, even the most innovative practices remain personal achievements that successors may ignore without penalty.

At its core, role expansion requires corporate autonomy: the legal and administrative authority for an institution to modify its own positions, expectations, and evaluative criteria. This autonomy allows internal actors—faculty, chairs, or corporate bodies—to redefine what a role is and what counts as competent performance. Without such autonomy, positions remain fixed by external statutes, charters, or endowments, and no amount of individual innovation can alter the institutional floor (Chamberlain 1994; Makdisi 1981).

## The Vesalius-Padua Transformation

The clearest demonstration of role expansion in operation is the transformation of the anatomy chair at the University of Padua under Andreas Vesalius. Before Vesalius, the anatomy chair was defined by Galenic textual commentary: the professor read authoritative texts aloud while a barber-surgeon performed the dissection. The authority of the professor derived from philological mastery, not anatomical practice. Dissection was a demonstration of what the texts already claimed, not an investigation of what bodies actually revealed (Klestinec 2011; Siraisi 1990).

Vesalius decisively redefined the role. He personally performed dissections rather than delegating to surgeons. He publicly corrected Galenic anatomy where observation contradicted textual authority. He replaced textual commentary with direct demonstration, making the body itself the source of anatomical knowledge. He embedded empirical anatomy into the pedagogical routine, requiring students to observe structures directly rather than memorizing descriptions. These changes are documented both in *De humani corporis fabrica* (Vesalius 1543) and in contemporary accounts of his teaching at Padua (Siraisi 2008, 1990; Klestinec 2011).

Significantly, Padua's corporate autonomy allowed this redefinition to become permanent. The university was legally empowered to modify its own positions; neither the Venetian Senate nor ecclesiastical authorities controlled the content of the anatomy chair (Biagioli 1993; Siraisi 1990). As a result, Vesalius's elevated standard became the new definition of the chair. His successors—Realdo Colombo, Gabriele Falloppia, Girolamo Fabricius ab Acquapendente, and eventually William Harvey—were required to meet this standard to occupy the role. The chair was no longer for Galenic commentary; it was for empirical anatomical demonstration (Klestinec 2011).

The mechanism is not psychological replication. Successors did not need Vesalius's drive; they needed only to satisfy the institutionally redefined expectations of the position. A candidate who offered to teach anatomy through Galenic commentary would not have been hired. The position itself now required empirical practice. Role expansion thus converts rare individual psychology into structural obligation.

## The Contrast Case: Islamic Madrasas

Islamic madrasas provide the clearest negative case. Their positions were defined by *waqf* endowment deeds, which legally fixed curriculum, teaching duties, and permissible texts. Because the *waqf* was a binding legal instrument interpreted through Islamic law, neither faculty nor administrators could modify a position's content. The endowment specified what was to be taught and how, and these specifications were enforceable in religious courts (Makdisi 1981; Chamberlain 1994).

Even highly innovative scholars—such as Ibn al-Haytham in optics or al-Tusi in astronomy—could not embed their elevated standards into institutional roles. Their achievements remained optional, adoptable by admirers but never enforceable as professional requirements. A madrasa professor could choose to teach Ibn al-Haytham's optical methods, but could equally choose not to. No institutional mechanism compelled adoption. When the individual scholar died, the elevated standard died with them unless a successor voluntarily chose to continue it. Without corporate autonomy, there was no pathway from individual innovation to institutional obligation (Sabra 1989; Ragep 2001).

The contrast isolates the mechanism:

- Padua possessed corporate autonomy → role expansion possible → standards embedded.

- Madrasas lacked corporate autonomy → role expansion impossible → standards remained personal.

### Why Role Expansion Matters

Role expansion solves the first institutional problem: how a single investigator's elevated practice becomes the minimum standard for all future practitioners. Without role expansion, drive produces brilliance but not transformation. With it, the institution itself becomes a conduit through which individual innovation is converted into mandatory competence.

But role expansion is necessary, not sufficient. The mechanism embeds standards, but it does not guarantee their persistence across generations. A single innovative incumbent can redefine a role, yet nothing prevents later holders from reverting to earlier adequacy levels unless the institution possesses a second mechanism: succession ratchets, which make elevated standards irreversible.

## 4.2. Succession Ratchets

Role expansion embeds elevated standards into the definition of a position, but does not guarantee that those standards will persist. A single innovative incumbent can redefine a role, yet nothing prevents later holders from reverting to earlier adequacy levels unless the institution possesses a second mechanism: succession ratchets. Succession ratchets make elevated standards irreversible by ensuring that each new appointment must meet or exceed the floor set by predecessors.

A succession ratchet requires three structural conditions:

1. **Competitive selection** — more qualified candidates than available positions.

2. **Incumbent control** — current officeholders evaluate and select successors.

3. **Professional judgment** — selection based on demonstrated competence rather than external criteria such as lineage, wealth, or patronage.

When these conditions align, the institution channels universal motivations—status, employment, prestige—toward maintaining elevated standards. Successors do not need to share the originating investigator's drive; they need only to want the position. The ratchet operates through institutional incentives, not psychological replication.

## The Padua Anatomy Chain

The anatomy chair at Padua provides the clearest demonstration of a functioning succession ratchet. After Vesalius redefined the role through direct anatomical demonstration, his successors—Realdo Colombo, Gabriele Falloppia, Girolamo Fabricius ab Acquapendente, and eventually William Harvey—were selected through competitive processes in which incumbents evaluated candidates based on anatomical competence (Klestinec, 2011; Siraisi, 1990, 2010).

Each successor was required to meet the elevated standard Vesalius had embedded into the position. The ratchet is visible in the historical record:

- **Colombo** (1544–1559) expanded Vesalius's corrections of Galen, introducing new anatomical findings from vivisection and further challenging textual authority. His *De re anatomica* (1559) shows continued commitment to empirical demonstration over textual commentary (Siraisi 2008).

- **Falloppia** (1551–1562) introduced new anatomical structures (including the fallopian tubes), refined dissection protocols, and extended comparative anatomy. His work assumed Vesalian standards as the baseline and built upon them (Klestinec, 2011).

- **Fabricius ab Acquapendente** (1565–1619) institutionalized comparative anatomy, built permanent anatomical theaters, and developed systematic embryological studies. His *De formato foetu* (1600) and *De venarum ostiolis* (1603) represent further elevation of empirical standards (Adelmann 1942; Klestinec 2011).

- **William Harvey**, trained under Fabricius (1600–1602), produced *De motu cordis* (1628), which depended entirely on the empirical anatomical culture Vesalius had initiated. Harvey's demonstration of circulation required systematic vivisection and quantitative measurement—methods that were obligatory at Padua but would have been unthinkable before Vesalius (Harvey 2016; French 1994).

At no point could a successor revert to pre-Vesalian textual commentary. The competitive structure of Padua's medical faculty, combined with incumbency control over appointments, made institutional regression impossible. A candidate who offered to teach anatomy through Galenic texts rather than dissection would not have been selected. The position itself enforced the elevated standard (Klestinec 2011; Siraisi 1990).

## The Savilian Professorships at Oxford

A similar ratchet operated in the Savilian mathematical chairs at Oxford. Henry Savile's founding statutes (1619) required incumbents to demonstrate mathematical competence and empowered current professors to evaluate successors (Feingold 1984). The result was a rising floor:

- **Henry Briggs** (Savilian Professor of Geometry, 1619–1630) established high standards in logarithmic and trigonometric computation. His *Arithmetica logarithmica* (1624) set new expectations for mathematical precision and utility (Havil 2014).

- **John Wallis** (Savilian Professor of Geometry, 1649–1703) expanded the mathematical curriculum, introduced analytic methods, and elevated standards for algebraic and infinitesimal techniques. His *Arithmetica infinitorum* (1656) assumed Briggs's computational standards and extended them into new domains (Stedall 2003).

- Later incumbents—including Edmond Halley and James Gregory—were required to match or exceed these standards to be considered credible candidates. The competitive evaluation process ensured that mathematical competence could not decline without disqualifying a candidate (Feingold 1984).

Again, the mechanism is institutional rather than psychological. Wallis did not need Briggs's investigative temperament; Wallis needed to satisfy the competitive expectations of the Savilian chair. The ratchet operated through selection criteria, not through psychological transmission.

## Contrast Case: Pre-Reformation English Universities

Before the Savilian statutes, mathematical positions at Oxford and Cambridge were often filled through patronage, clerical preferment, or seniority rather than demonstrated competence. Positions were rewards for loyalty or service, not incentives for elevated practice. Because selection was not competitive and incumbents did not control succession, there was no mechanism to prevent regression. A highly competent mathematician might hold a chair, but the next appointee could be mediocre without institutional penalty (Feingold 1984).

The contrast isolates the mechanism:

- Savilian chairs possessed competitive selection + incumbent control → succession ratchet operated → standards rose irreversibly.

- Pre-reformation chairs lacked competitive selection + incumbent control → no ratchet → standards fluctuated with individual competence.

### Why Succession Ratchets Matter

Succession ratchets solve the second institutional problem: how elevated standards persist across generations even when the originating investigator is gone. This mechanism explains why the Scientific Revolution persisted in Europe but dissipated in other civilizations. Islamic natural philosophy during the Golden Age and Chinese astronomy during the Song Dynasty both produced brilliant individual investigators, but their institutional structures differed significantly from those of European universities. Chinese astronomy functioned as a state bureaucracy: when dynasties changed, or imperial patronage shifted, institutional memory could be disrupted. Islamic astronomy developed sophisticated traditions (notably the Maragha School), but these remained dependent on ephemeral princely patronage rather than durable, autonomous corporate structures (Sivin 1995; Ragep 2001).

Succession ratchets ensure that:

- standards cannot be lowered without professional penalty,
- regression is institutionally blocked, and
- rare individual innovations become permanent features of the discipline.

Without succession ratchets, role expansion produces only temporary elevation. A single brilliant incumbent raises the bar, but the next appointment reverts to earlier norms. With succession ratchets, elevated standards become irreversible. Each generation must meet or exceed the floor set by predecessors. The floor is the critical difference between episodic brilliance and cumulative transformation.

Succession ratchets, however, do not determine *where* rising standards will occur. That requires the third mechanism: domain channeling, which directs institutional attention and prestige toward particular fields.

### 4.3. Domain Channeling: Why These Mechanisms Operated in Natural Philosophy

Role expansion and succession ratchets explain how elevated standards can be embedded and preserved within institutions. But these mechanisms do not explain *where* they operate. The same institutional machinery can raise standards in natural philosophy, classical philology, theology, or law. The third mechanism—domain channeling—determines which domain benefits from rising standards.

Domain channeling operates through two complementary factors. Prestige gradients direct institutional energy, resources, and competitive attention toward particular fields. These gradients arise from examination systems, patronage priorities, state needs, and cultural hierarchies of knowledge. Domain affordances determine whether a field can productively absorb channeled energy. A domain may receive institutional attention but lack the internal structure to convert that attention into rising investigative standards.

Natural philosophy became the site of the Scientific Revolution because both conditions held: prestige flowed toward it (channeling), and it possessed the structural features to absorb that energy (affordances).

### 4.3.1. Prestige Gradients

Prestige gradients are not psychological; they are structural. They arise from institutional configurations that reward expertise in particular domains with status, employment, and resources. When prestige is concentrated in a domain, institutions channel their role-expansion and succession-ratchet mechanisms into that field, producing rising standards there rather than elsewhere.

*The Chinese Examination System: Channeling Toward the Classics*

Imperial China provides the clearest demonstration of domain channeling through prestige gradients. The *keju* civil service examinations—the primary route to elite status—tested mastery of the Confucian classics, not natural philosophy. As Benjamin Elman has shown, the examination system created a powerful prestige gradient: intellectual ambition, institutional resources, and elite competition were all directed toward classical scholarship (Elman 1991, 2000). Success in the examinations determined access to government positions, social mobility, and cultural recognition. Natural philosophy, by contrast, offered no comparable rewards.

The result was a highly sophisticated philological tradition, complete with role expansion (textual criticism became mandatory for elite scholars) and succession ratchets (competitive examinations enforced rising standards of classical mastery).

Investigators such as Gu Yanwu (1613–1682) exemplify this pattern: his methodological rigor in evidential scholarship (*kaozheng*) displayed clear investigative drive—systematic textual criticism that exceeded practical utility, sustained engagement across decades of phonological and geographical research, and severe framing of philological errors as fundamental inadequacies (Chow 1994; Elman 2000). But because the *keju* system did not reward natural-philosophical innovation, the institutional machinery for rising standards was channeled toward classical studies rather than natural inquiry (Chow 1994; Elman 2000).

China, therefore, possessed the institutional mechanisms for rising standards—the *kaozheng* movement created peer-review-like standards, specialized scholarly networks, and succession ratchets where students tried to outdo teachers in philological precision—but they operated in the domain of classical scholarship rather than natural philosophy (Elman 2005).

*Islamic Civilization: Channeling Toward Theology and Law*

A similar pattern appears in Islamic civilization. The highest prestige domains were *fiqh* (jurisprudence) and *kalam* (theology), not natural philosophy. Madrasas were endowed through *waqf* deeds that specified curricula centered on law and theology, and elite status was tied to mastery of these fields (Chamberlain 1994; Makdisi 1981). Scholars who achieved eminence in jurisprudence or theology gained institutional positions, social recognition, and influence. Natural philosophy, while respected, did not confer comparable status.

As a result, institutional energy—appointments, patronage, scholarly competition—was channeled toward legal and theological expertise. This channeling explains why even brilliant natural-philosophical investigators such as Ibn al-Haytham (965–1040) did not generate durable traditions. His optical work displayed investigative drive: recursive puzzle generation, experimental rigor, precision beyond utility, and severity of framing—his *Doubts on Ptolemy* subjected inherited astronomy to systematic critique (Sabra 1989). His *Kitāb al-Manāẓir* (*Book of Optics*) introduced systematic experimental methods and challenged inherited Greek optical theories (Sabra 1989). But the institutional mechanisms capable of embedding and ratcheting standards operated in theology and law, not in natural philosophy. Ibn al-Haytham's achievements were admired and studied, but his methods did not become mandatory professional standards because prestige gradients directed institutional machinery elsewhere (Ragep 2001; Sabra 1989).

### Medieval Europe: Channeling Toward Theology

Medieval European universities also illustrate domain channeling through prestige gradients. Their highest-status faculties were theology, law, and medicine; natural philosophy was subordinate and largely preparatory. The university curriculum was structured hierarchically: students studied arts (including natural philosophy) as preparation for advanced study in theology, law, or medicine (Weisheipl 1964; Grant 1996).

Even where role expansion and succession ratchets existed—especially in theology— these mechanisms elevated standards within theological disputation rather than natural inquiry. The institutional structure thus produced rising standards, but in a domain that did not generate the recursive, instrument-driven tensions characteristic of the Scientific Revolution. Theological questions could be pursued with great sophistication, but they did not progress through more refined measurement or experimental intervention. The prestige gradient directed institutional energy toward theology (Leff 1975; Grant 1996).

### Early Modern Europe: Channeling Toward Natural Philosophy

The Scientific Revolution occurred only when prestige gradients shifted. In sixteenth- and seventeenth-century Europe, several forces redirected institutional attention toward natural philosophy:

- **State patronage for astronomy**: Calendrical reform (the Gregorian calendar, 1582), navigation (longitude determination, shipbuilding), and artillery (ballistics, fortification engineering) created practical demand for astronomical and mathematical expertise. States employed mathematicians and astronomers as technical advisors (Biagioli 1993; Bennett 1987).

- **Courtly interest in mathematical and mechanical expertise**: European courts valued mechanical devices, mathematical instruments, and natural-philosophical demonstrations as markers of sophistication. Court mathematicians gained status through spectacular innovations—Galileo's telescope, Santorio's instruments, Kepler's astronomical tables. Patronage followed innovation (Findlen 1994; Biagioli 1993).

- **The rise of instrument-based knowledge**: Telescopes, microscopes, air pumps, and precision clocks created new investigative opportunities and new forms of demonstration. Instrument-makers and natural philosophers formed collaborative networks that elevated the status of empirical investigation (Bennett 1987; Shapin and Schaffer 2011; Van Helden 1977).

- **The loosening of university control over natural-philosophical inquiry**: Natural philosophy increasingly occurred outside universities—in courts, workshops, correspondence networks, and private academies. This institutional looseness made it easier for innovative practices to gain traction without ecclesiastical or curricular gatekeeping (Biagioli 1993; Shapin 1996).

These shifts created a new prestige gradient: natural philosophy became a domain in which ambitious investigators could gain status, patronage, and institutional recognition. Once prestige flowed into natural philosophy, role expansion and succession ratchets began operating there, producing rising standards in observation, measurement, and mathematical modeling.

The prestige gradient thus explains *why* institutional mechanisms operated in natural philosophy rather than in theology, law, or classical scholarship. Prestige directed the machinery; the machinery elevated standards.

### 4.3.2. Structural Features of Natural Philosophy

Prestige gradients explain why institutional energy flowed *toward* natural philosophy, but structural features explain why that energy *could be absorbed*. A domain may receive institutional attention but lack the internal structure to convert that attention into rising investigative standards. Natural philosophy possessed a distinctive combination of structural features that made it unusually capable of sustaining the institutional mechanisms described above.

*1. High-Resolution Tensions*

Natural philosophy contained tensions that were both conceptually deep and empirically tractable. Problems in motion, optics, astronomy, and mechanics could be interrogated at increasingly fine levels of precision. Ptolemaic astronomy delivered predictions within observational tolerances, but discrepancies accumulated as instruments improved. Aristotelian mechanics explained everyday phenomena qualitatively, but failed under systematic quantification. These tensions were therefore well suited to activation: they generated intolerable gaps that drove investigators into recursive interrogation (Koyré 1957; Swerdlow and Neugebauer 2012). Anomalies in natural philosophy were material rather than interpretive. A disputed statute could be reinterpreted; a theological tension could be resolved through doctrinal distinction; a textual contradiction could be harmonized through exegesis. But Mars's orbit could not be hermeneutically adjusted to fit circular motion, and Jupiter's moons could not be exegetically reconciled with geocentrism. An 8-minute orbital discrepancy or four new celestial bodies demanded systematic reconstruction rather than interpretive refinement. As Kuhn observed, certain anomalies resist accommodation within existing frameworks and demand conceptual restructuring (Kuhn 2012). To use a contemporary analogy, natural philosophy presented hardware conflicts that crashed existing systems, while law and theology presented software patches that could be installed without rebuilding foundations.

This distinction explains why natural philosophy was structurally more receptive to activation than other domains. Legal and theological tensions, however deep, rarely forced investigators into the kind of recursive, precision-driven interrogation that generates psychological intolerance. They could be indefinitely negotiated, deferred, or accommodated. The material anomalies of natural philosophy, by contrast, became more rather than less acute as investigative precision increased—exactly the pattern that triggers activation.

*2. Instrumental Fruitfulness*

Natural philosophy was uniquely positioned to benefit from new instruments—telescopes, microscopes, air pumps, pendulum clocks—that amplified small investigative steps into cascades of new puzzles (Bennett 1987; Shapin and Schaffer 2011; Van Helden 1977). These instruments created the material scaffolding necessary for capture: they stored partial results, generated new anomalies, and required continual refinement.

The telescope exemplifies this instrumental fruitfulness. Galileo's early observations of Jupiter's moons generated immediate puzzles about orbital mechanics. Improved magnification revealed the phases of Venus, sunspots, and lunar topography—each requiring further investigation. The instrument itself became a source of recursive puzzle generation. Other domains lacked comparable material infrastructures. Theology had no instruments to amplify textual tensions; law had no devices to generate cascades of interpretive puzzles.

### 3. Representational Plasticity

Natural philosophy's representational systems—geometrical diagrams, algebraic formulations, mechanical analogies—were unusually plastic and could be reconfigured to express new problem-structures, enabling the reformatting necessary for ratchet (Nersessian 1999; Dear 2019; Gaukroger 2006).

Kepler could reformulate planetary motion from circles and epicycles into ellipses and area laws. Galileo could reformulate motion from qualitative tendencies into quantitative kinematic relations. Descartes could reformulate geometry into algebra. Each reformulation created a new problem space that successors entered into naturally. By contrast, the representational systems of law, theology, and philology were more rigid: they were anchored in inherited textual corpora that constrained reformulation. Legal reasoning depended on statutory interpretation; theological reasoning depended on scriptural and patristic authority. These systems resisted the kind of radical restructuring that natural philosophy underwent.

### 4. Weak Institutional Gatekeeping

Natural philosophy lacked the strong institutional gatekeeping found in law, medicine, and theology. Universities controlled professional entry into those fields, enforcing curricular orthodoxy and limiting the uptake of radical innovations. To practice law or medicine required university credentials and guild membership. To teach theology required ecclesiastical approval and doctrinal conformity (Grant 1996; Siraisi 1990).

Natural philosophy, by contrast, was dispersed across courts, workshops, correspondence networks, and private academies (Biagioli 1993; Shapin 1996). This institutional looseness made it easier for captured inquiries to circulate and for ratcheted problem-structures to gain traction. Galileo could publish telescopic discoveries without university approval. Kepler could reformulate planetary theory outside traditional astronomical faculties. Boyle could conduct experimental demonstrations in private laboratories and through the Royal Society. The absence of gatekeeping allowed innovative practices to bypass institutional conservatism.

### 5. Cross-Domain Utility

Finally, natural philosophy had unusually broad utility. Innovations in mechanics, optics, and astronomy had immediate applications in navigation, artillery, surveying, and instrument-making (Bennett 1987; Long 2011; Smith 2004). These applications created additional channels of capture and anchoring: artisans, patrons, and state actors became entangled in the investigative trajectories.

Navigation required accurate astronomical tables for determining longitude. Artillery required ballistic calculations. Surveying requires geometric methods and precision instruments. These practical demands created constituencies beyond academic natural philosophers—shipbuilders, military engineers, instrument-makers, state administrators—who had material interests in natural-philosophical innovations. Other domains lacked comparable cross-domain leverage. Theological disputation had no naval applications; philological method had no military utility.

*Why Domain Affordances Matter*

Taken together, these affordances explain why natural philosophy could productively absorb the institutional energy directed toward it by prestige gradients. Natural philosophy possessed:

- tensions that could be recursively refined (high-resolution),
- instruments that amplified investigation (instrumental fruitfulness),
- representational systems that could be radically reformatted (plasticity),
- institutional structures that did not block innovation (weak gatekeeping), and
- practical applications that created external constituencies (cross-domain utility).

Other domains received institutional attention—theology in medieval Europe, law and theology in Islamic civilization, classical scholarship in China—but lacked the full set of affordances. These domains could not convert prestige into rising investigative standards because their internal structure resisted the recursive, instrument-driven, problem-reformatting dynamics that natural philosophy enabled.

The Scientific Revolution, therefore, occurred in natural philosophy not because it was uniquely rational or conceptually privileged, but because it possessed the structural affordances that allowed activated inquiries to become captured and ratcheted into durable traditions. Domain channeling—the combination of prestige gradients and domain affordances—directed the full machinery of role expansion and succession ratchets toward natural philosophy, enabling the transformation of investigative drive into a cumulative research tradition.

# 5. Comparative Configurations: Testing the Alignment

The preceding sections have identified six components of the trigger architecture: activation, capture, and externalization at the individual level; role expansion, succession ratchets, and domain channeling at the institutional level. These components are operationalizable mechanisms inferred from behavioral evidence and institutional structure.

This section tests the framework through systematic variation. By examining historical configurations in which different subsets of these components aligned, we isolate their individual and joint causal contributions. The method is controlled comparison: when some mechanisms are present, and others are absent, the resulting differences in outcomes reveal which components are necessary and which combinations are sufficient.

The analysis proceeds in two stages. Partial alignments show what happens when some but not all mechanisms are present. Full alignments show what happens when all components converge. Together, these cases demonstrate that the Scientific Revolution emerged not from unique European capacities but from a specific alignment of mechanisms at particular sites.

**Partial Alignments**

**Drive Without Institutional Embedding: Copernicus and Ibn al-Haytham**

Nicolaus Copernicus (1473–1543) displayed clear activation: severity of framing (calling Ptolemaic astronomy "a monster"), precision beyond utility (theoretical coherence over predictive accuracy), and sustained engagement (over thirty years of revision). Yet *De revolutionibus* (1543) remained adoptable but not obligatory. Universities continued teaching Ptolemaic astronomy; no institutional penalty accompanied rejection of heliocentrism. Without mechanisms to embed heliocentric standards into positions or enforce them across generations, Copernican astronomy influenced later investigators voluntarily but never became a professional requirement (Westman 2011).

Similarly, Ibn al-Haytham (965–1040) demonstrated the mechanism of activation in optics through rigorous experimentation, precision beyond existing standards, the severity of framing, and sustained engagement across his *Kitāb al-Manāẓir* (Sabra 1989). His methods were studied and extended by later scholars—Kamāl al-Dīn al-Fārisī, Quṭb al-Dīn al-Shīrāzī—but Islamic madrasas lacked the corporate autonomy necessary for role expansion. Waqf endowments legally fixed curricula, preventing faculty from embedding elevated standards into positions. Without institutional mechanisms to make his standards mandatory, these methods remained optional practices that dissipated when not voluntarily adopted (Ragep 2001).

Islamic astronomy did produce institutional succession—notably the Maragha School, from al-Ṭūsī to al-Shāṭir, which generated increasingly sophisticated planetary models. However, these remained dependent on princely patronage rather than embedded in the educational core of the madrasa system (Ragep 2001).

**Result**: Drive produces influential achievements that do not become professional obligations. Activation is necessary but not sufficient.

**Externalization Without Institutional Reinforcement: Santorio**

Santorio Santorio's metabolic work at Padua demonstrates a critical intermediate configuration: activation and externalization without institutional reinforcement. Santorio occupied the first chair of theoretical medicine at Padua from 1611 to 1624, during which he developed quantitative methods for measuring fever, pulse, and insensible perspiration. His work displayed clear activation markers: he pursued precision beyond any therapeutic utility (quantitative temperature measurement served no immediate clinical purpose), sustained engagement across three decades, and framed the absence of measurement as a fundamental inadequacy of traditional Galenic medicine. His *Ars de* *statica medicina* (1614) externalized these methods through standardized protocols, calibrated instruments, and systematic procedures that others could adopt (Bigotti 2020).

The work achieved remarkable individual-level externalization. *Medicina statica* saw more than 55 reprints over 150 years, becoming, as Barry and Bigotti call it, "a fundamental textbook in European medical practice." Santorio's name became synonymous with the weighing chair (the "Sanctorian chair") and with the measurement of insensible perspiration ("Sanctorian perspiration"). His methods "remained a living part of scientific and medical theory and practice until at least the time of Lavoisier," and "successive generations of scholars were inspired to conduct their own programmes of experimentation and theorising by following his lead" (Bigotti 2020). By any measure of individual-level externalization—circulation, adoption, influence—Santorio's work succeeded.

Yet this externalization was not reinforced by institutional mechanisms. When Santorio resigned in 1624 due to political conflicts, he was replaced by Pompeo Caimo, described by Bigotti as "less talented but openly pro-papal" (Bigotti 2020). The appointment was political rather than competitive; there is no evidence that candidates were evaluated on the basis of quantitative methodological competence. Unlike the anatomy chair, where each successor was required to match or exceed Vesalian standards of empirical demonstration, the first chair of medicine did not embed Santorio's quantitative methods as mandatory requirements. His methods remained available—widely circulated, admired, adopted by some—but never obligatory. Competitive selection did not enforce them; succession ratchets did not preserve them.

The contrast with Vesalian anatomy is instructive. Both programs operated at the same institution during overlapping periods. Both achieved individual-level externalization. But only anatomy possessed the institutional mechanisms to convert externalized methods into permanent professional requirements. Santorio's case, therefore, isolates the necessity of institutional reinforcement: externalization creates the possibility of persistence, but only succession ratchets—competitive selection enforcing rising standards—make that persistence durable and irreversible. Without them, even widely influential innovations remain optional rather than obligatory.

**Institutional Mechanisms Operating in Other Domains: Medieval Europe**

Medieval European universities possessed functional role expansion and succession ratchets in theology. The Paris theology faculty could redefine positions without external authorization, competitive disputations enforced rising standards, and successive generations—Thomas Aquinas, Duns Scotus, William of Ockham—elevated standards of argumentative rigor and logical precision (Leff 1975). The mechanisms operated effectively, but in theology rather than natural philosophy. Natural philosophy remained subordinate and preparatory; prestige, competitive pressure, and institutional energy flowed toward theological disputation (Grant 1996).

**Result**: Institutional mechanisms elevate standards where prestige directs them. Role expansion and succession ratchets are insufficient without domain channeling toward natural philosophy.

**Channeling Toward Non-Natural-Philosophical Domains: China and Islam**

Imperial China's keju examination system—the primary route to government positions and elite status—tested mastery of the Confucian classics, not natural philosophy. The system created a powerful prestige gradient toward classical scholarship. Scholars like Gu Yanwu (1613–1682) displayed investigative drive in evidential scholarship (kaozheng): precision beyond utility, sustained engagement, severity of framing. The system possessed effective mechanisms for raising standards—scholarly reputations could redefine what counted as adequate classical scholarship, and competitive examinations enforced rising rigor. But because examinations rewarded classical mastery rather than natural-philosophical innovation, these mechanisms elevated standards in philology, not natural investigation (Elman 2000).

Islamic civilization similarly channeled prestige toward fiqh (jurisprudence) and kalam (theology). Madrasas were endowed specifically for legal and theological instruction; scholars who achieved eminence in these fields gained institutional positions, patronage, and influence. Even brilliant natural philosophers like Ibn al-Haytham operated outside the main channels of institutional prestige. The mechanisms capable of embedding and preserving elevated standards—teaching positions, scholarly networks, competitive recognition—operated in law and theology, not natural philosophy (Makdisi 1981; Chamberlain 1994).

**Result**: Prestige gradients determine where institutional mechanisms operate. Without channeling toward natural philosophy, drive and institutional capacity produce sophisticated traditions elsewhere.

**Drive Without Institutional Continuity: Ancient Greece**

Ancient Greece produced investigators with extraordinary drive—Archimedes, Hipparchus, Ptolemy, Eratosthenes—but lacked corporate bodies with authority to redefine positions, competitive succession mechanisms, or institutional continuity across generations. Greek natural philosophy occurred in informal contexts: philosophical schools (Lyceum, Academy), royal patronage (Alexandria, Syracuse), and individual investigation. These contexts could support brilliant individuals, but when an investigator died, elevated practices often died with them unless a dedicated student voluntarily chose to continue the work. Archimedes' methods were transmitted episodically to later scholars—Diocles, Eutocius, medieval Arabic mathematicians—but no institution enforced Archimedean standards as mandatory for practitioners (Lloyd 1973; Netz 2003).

The contrast with Galileo is instructive: Galileo's Archimedean-style mathematical methods became embedded in institutional positions and preserved through succession mechanisms. Archimedes did not. The difference lies not in individual achievement but in institutional scaffolding.

**Result**: Drive without institutional continuity produces episodic breakthroughs without cumulative accumulation. Activation and externalization are insufficient without institutional embedding and preservation.

**Full Alignments**

**Padua-Venice: Anatomy and Physiology**

The Padua anatomy sequence demonstrates complete alignment. Andreas Vesalius displayed activation: severity of framing (anatomy had been "miserably destroyed" by textual reliance), precision beyond utility (Galenic anatomy was adequate for medical practice; Vesalius pursued epistemic accuracy), and sustained engagement. Padua's corporate autonomy—the university's legal authority to modify positions without Venetian or ecclesiastical approval—enabled role expansion: Vesalius's redefinition of the anatomy chair from textual commentary to empirical demonstration became institutionally binding. Competitive appointments created succession ratchets: Realdo Colombo, Gabriele Falloppia, Girolamo Fabricius ab Acquapendente, and William Harvey were required to meet and exceed Vesalian standards to occupy the chair. Venetian medical prestige provided domain channeling: institutional energy flowed toward empirical anatomy rather than other fields.

The result was a cumulative research tradition lasting over a century. Each successor elevated standards further: Colombo introduced vivisection, Falloppia refined dissection protocols, Fabricius institutionalized comparative anatomy and built permanent theaters, and Harvey demonstrated circulation through systematic vivisection and quantitative measurement. Harvey's achievement (1628) depended entirely on the empirical culture Vesalius initiated. The configuration transformed individual innovation into collective obligation (Klestinec 2011; Siraisi 2008, 1990).

**Oxford-London: Mathematics and Experimental Philosophy**

The Oxford-London corridor shows a parallel alignment. Henry Savile's founding statutes (1619) for the Savilian professorships at Oxford created role expansion (positions could be redefined through demonstrated mathematical competence) and succession ratchets (current professors evaluated successors competitively). Henry Briggs established high standards in logarithmic computation; John Wallis was required to match these standards to be appointed and then to extend them into infinitesimal methods; later incumbents—Edmond Halley and David Gregory—faced progressively higher floors (Feingold 1984).

In London, the Royal Society (chartered 1660) provided institutional anchoring for experimental work. Robert Boyle demonstrated activation in his air-pump experiments; his methods were externalized as experimental recipes in *New Experiments Physico-Mechanical* (1660). Robert Hooke, as Curator of Experiments, was institutionally required to produce weekly demonstrations, creating continuous innovation. State patronage for navigation, artillery, and astronomy channeled prestige toward natural philosophy through positions at the Royal Observatory at Greenwich, the Admiralty, and the Mint (Shapin and Schaffer 2011; Hunter 1998).

Together, these institutions created an ecosystem that amplified individual drive. Isaac Newton's *Principia Mathematica* (1687) emerged from this configuration: Savilian-style mathematical rigor, Royal Society experimental culture, and state-supported astronomical observation. The work became canonical not just because of its brilliance, but because institutional mechanisms embedded its standards as professional requirements. Later mathematicians and natural philosophers were trained in Newtonian methods; deviations from Newtonian standards required justification (Westfall 1983; Guicciardini 1999).

**What the Configurations Demonstrate**

This systematic comparison isolates the causal architecture:

- **Activation alone** → influential but optional innovations (Copernicus, Ibn al-Haytham)
- **Institutional mechanisms alone** → rising standards in other domains (medieval theology)
- **Channeling elsewhere** → institutional energy directed away from natural philosophy (China, Islam)
- **Drive without continuity** → episodic achievements without accumulation (ancient Greece)
- **Full alignment** → durable, cumulative research traditions (Padua-Venice, Oxford-London)

The comparison demonstrates that:

1. **All six components are necessary**: No configuration lacking any component produced the Scientific Revolution's distinctive outcome
2. **No single component is sufficient**: Drive, institutions, or channeling alone did not produce cumulative transformation
3. **Alignment is localized**: The Scientific Revolution occurred at Padua-Venice and Oxford-London, not uniformly across Europe
4. **The configuration is contingent**: Other civilizations possessed drive, institutional mechanisms, and sophisticated traditions, but configured them differently

The Scientific Revolution emerged not from unique European capacities but from a specific convergence of psychological and institutional mechanisms at particular sites. Understanding it requires analyzing not the presence versus absence of components but patterns of alignment.

# 6. Conclusion

This paper set out to resolve three puzzles about the Scientific Revolution: what triggered the initial escalation of inherited tensions (activation), what made early investigative efforts durable rather than ephemeral (durability), and why natural philosophy became the locus of transformation in Europe rather than theology, law, or classical scholarship (direction). The trigger architecture provides answers.

The activation puzzle is resolved by identifying a psychological threshold at which inherited tensions become intolerable and demand resolution. While dispositional factors such as intellectual persistence may predispose some investigators toward intensive interrogation, disposition alone cannot explain within-individual selectivity. The threshold emerges through intensive interrogation that creates gaps others do not perceive, operating as a context-dependent state rather than a stable trait carried uniformly across domains. Galileo's selective activation across domains—escalating telescopic astronomy while treating thermometry as a qualitative demonstration—confirms that activation operates as a content-dependent mechanism emerging from the interaction among investigator characteristics, problem structure, and interrogation depth, rather than as a stable personality trait manifesting uniformly across contexts. The contrast between investigators who shared external conditions but produced divergent outcomes (Galileo vs. Harriot, Galileo vs. Santorio) isolates this threshold as a distinct causal component.

The durability puzzle is resolved by capture and externalization at the individual level, reinforced by role expansion and succession ratchets at the institutional level. Galileo's telescopic work was captured through recursive puzzle generation (each discovery opened new questions), material inertia (apparatus investment), and social tethering (correspondence networks, controversy). His methods were externalized through protocols others could adopt. But durability required institutional reinforcement: Padua's corporate autonomy allowed Vesalius to embed elevated anatomical standards into positions; competitive succession prevented regression across generations. Santorio's metabolic work, by contrast, was activated and externalized but not institutionally reinforced—the trajectory dissipated within a generation.

The direction puzzle is resolved by domain channeling—the combination of prestige gradients and structural affordances that directed institutional energy toward natural philosophy. Medieval universities channeled the same mechanisms (role expansion, succession ratchets) toward theology. Imperial China channeled them toward classical scholarship through the *keju* examination system. Islamic civilization channeled them toward jurisprudence and theology. Early modern Europe redirected prestige toward natural philosophy through state patronage (navigation, artillery), courtly interest (mechanical instruments), and the loosening of university control. Natural philosophy could absorb this energy because it possessed high-resolution empirical tensions, material amplifiability through instruments, representational plasticity, weak institutional gatekeeping, and cross-domain utility. Metaphorically, natural philosophy presented "hardware conflicts"—material anomalies that could not be resolved through textual reinterpretation—while law and theology offered "software patches" that could be installed through hermeneutic refinement.

**A Contingent Alignment, Not European Destiny**

The Scientific Revolution emerged from a contingent alignment of psychological and institutional mechanisms at particular sites. Nothing in this account implies European exceptionalism or teleological inevitability. Other civilizations possessed investigative drive: Ibn al-Haytham in optics, Gu Yanwu in evidential scholarship, Archimedes in mathematics. They possessed institutional mechanisms: Chinese examinations created succession ratchets, Islamic scholarly networks transmitted methods, and Greek philosophical schools supported inquiry. What they lacked was not capacity but alignment.

The distinctiveness of early modern Europe lay in specific, contingent conditions. Padua's corporate autonomy resulted from Venetian governance arrangements, not inherent European rationality. Oxford's competitive mathematical chairs reflected particular patronage priorities rather than predetermined Western progress. The loosening of ecclesiastical control over natural philosophy emerged from Reformation conflicts and jurisdictional fragmentation, not from a teleological march toward secularization. Had Venetian governance been different, had English patronage followed other priorities, had ecclesiastical control remained tighter, the alignment might not have occurred—or might have occurred elsewhere, at other times, under different conditions.

The alignment was localized, not diffuse. It crystallized at Padua-Venice (anatomy and physiology) and Oxford-London (mathematics, astronomy, experimental philosophy) during the late sixteenth and seventeenth centuries. These were not the only sites of natural-philosophical innovation in Europe, but they were the sites where all components converged: investigative drive met institutional mechanisms capable of embedding, preserving, and directing elevated standards toward natural philosophy. The Scientific Revolution was not a pan-European phenomenon but a specific configuration that emerged where particular conditions aligned.

**Generalizable Framework for Scientific Change**

Although this account has focused on the Scientific Revolution, the architecture it identifies is not confined to early modern Europe. The mechanisms—activation, capture, externalization, role expansion, succession ratchets, domain channeling—provide a general framework for analyzing episodes of scientific change, including those often described as paradigm shifts.

The Chemical Revolution shows similar patterns. Lavoisier's intolerance of phlogiston theory reflects activation: the severity of framing (calling it "absurd"), the precision beyond utility (quantitative measurements exceeding practical chemistry), and sustained engagement (decades of systematic experimentation). His oxygen theory was captured through apparatus standardization (gasometers, balances), recursive puzzle generation (combustion → respiration → acidity), and social networks (correspondence, académie demonstrations). Externalization occurred through experimental protocols and nomenclature reform. Institutional mechanisms then embedded the new standards: French educational reforms after 1789 expanded roles (chemistry chairs redefined around Lavoisian methods), competitive examinations created succession ratchets, and state support for technical education channeled prestige toward experimental chemistry.

The Darwinian case shows parallel dynamics. Darwin's activation is evident in his decades of private investigation beyond institutional demands, his obsessive precision in documenting variation, and his framing natural theology as fundamentally inadequate. Natural selection was captured through extensive empirical scaffolding (*Origin* relied on twenty years of data accumulation) and social networks (Hooker, Huxley, Lyell). Externalization occurred through the conceptual mechanism of natural selection, becoming a research program others could pursue without sharing Darwin's theological concerns. Institutional embedding followed: university positions in biology were redefined around evolutionary biology, competitive publication enforced rising evidential standards, and prestige flowed toward evolutionary research through state funding and professional societies.

These examples suggest that the trigger architecture identifies a generalizable mechanism-level substrate beneath diverse episodes of scientific change. Whenever investigative norms shift—from qualitative to quantitative, observational to experimental, descriptive to theoretical—similar processes may be at work: psychological thresholds crossed, early innovations captured, methods externalized, and institutional mechanisms embedding new standards as obligatory rather than optional.

**The Emergence of Modern Science**

The Scientific Revolution transformed the landscape of inquiry not through sudden conceptual rupture but through the accumulation of elevated standards that became irreversible. Individual compulsion was converted into collective obligation; episodic brilliance into durable traditions; and optional innovations into professional requirements. This transformation occurred through the alignment of psychological and institutional mechanisms at specific sites where corporate autonomy, competitive succession, and prestige gradients converged with investigative practices.

Understanding the Scientific Revolution in these terms shifts the explanatory focus. The question is not why Europe "succeeded" while other civilizations "failed," but how different historical settings combined the same components in different configurations. The Scientific Revolution was one possible outcome within a broader space of alignments—distinctive not because its components were unique, but because their convergence was. By identifying the mechanisms that link psychological thresholds to institutional structures, this account shows how investigative practices become stabilized, how standards rise irreversibly, and how domains shift. The emergence of modern science was neither accidental nor predetermined, but the product of a specific alignment that, once established, proved remarkably durable and productive.